\documentclass[conference]{IEEEtran}
\IEEEoverridecommandlockouts

\usepackage{cite}
\usepackage{amsmath,amssymb,amsfonts}
\usepackage{algorithmic}
\usepackage{graphicx}
\usepackage{textcomp}
\usepackage{xcolor}

\usepackage[colorlinks=true, allcolors=blue]{hyperref}
\def\BibTeX{{\rm B\kern-.05em{\sc i\kern-.025em b}\kern-.08em
    T\kern-.1667em\lower.7ex\hbox{E}\kern-.125emX}}

\usepackage{etoolbox}
\makeatletter
\patchcmd{\@maketitle}
  {\addvspace{0.5\baselineskip}\egroup}
  {\addvspace{-1.5\baselineskip}\egroup}
  {}
  {}
\makeatother

\renewcommand{\tablename}{ }
    
\begin{document}

\title{Near-term Application Engineering Challenges in Emerging Superconducting Qudit Processors}

\author{\IEEEauthorblockN{Davide Venturelli}
\IEEEauthorblockA{\textit{SQMS, NASA QuAIL \& USRA} \\
Mountain View, California, USA \\
dventurelli@usra.edu}
\and
\IEEEauthorblockN{Erik Gustafson}
\IEEEauthorblockA{\textit{SQMS, NASA QuAIL \& USRA} \\
Mountain View, California, USA \\
egustafson@usra.edu}
\and
\IEEEauthorblockN{Doga Kurkcuoglu}
\IEEEauthorblockA{\textit{SQMS, Fermilab} \\
Batavia, Illinois, USA \\
dogak@fnal.gov}
\and
\IEEEauthorblockN{Silvia Zorzetti}
\IEEEauthorblockA{\textit{SQMS, Fermilab} \\
Batavia, Illinois, USA \\
zorzetti@fnal.gov}
}
\maketitle
\vspace{-10pt}
\begin{abstract}
We review the prospects to build quantum processors based on superconducting transmons and radiofrequency cavities for testing applications in the NISQ era. We identify engineering opportunities and challenges for implementation of algorithms in simulation, combinatorial optimization, and quantum machine learning in qudit-based quantum computers. 
\end{abstract}

\vspace{-5pt}
\section{Introduction \& Previous Work}
Quantum computing with qu\textbf{d}its – quantum systems of dimension $d$\,$>$2– could enhance computational capacity beyond qu\textbf{b}it systems. 
The larger Hilbert space in qudits enables differently encoded quantum algorithms and simulations, potentially bringing resource utilization advantages with respect to qubits~\cite{wang2020qudits}.
However, solid state multi-level systems, are generally more susceptible to cross-talk, leakage, and decoherence, posing control challenges. 
Superconducting circuits are one of the most promising and studied platforms for quantum computing.
In superconducting circuits, the transmon device serves as a natural qudit, with multiple accessible energy levels in a single physical element. 
In circuit quantum electrodynamics (cQED) systems, the quantum processing unit (QPU) is achieved by coupling a superconducting transmon to a long-lived cavity oscillator, thus the multilevel cavity states can also be utilized as qudit-based QPU. In cQED, microwave cavity modes can be coplanar with the transmon (2D), or the electromagnetic mode in a 3D cavity can be coupled to a transmon.  
3D superconducting radiofrequency (SRF) cavities have redefined achievable quantum coherence in superconducting platforms, since they support electromagnetic modes with exceptionally long lifetimes, far exceeding those of conventional resonators or qubits.  These 3D cavities act as quantum memories for \emph{qumodes} or \emph{bosonic qudits}, offering a large Hilbert space in which quantum information can be stored and manipulated. A single cavity mode can serve as a $d$-level system with an effectively large $d$, enabling encoding of logical qudits or error-correctable bosonic states within the oscillator subspace.

In this study, we illustrate capabilities of these emerging architectures by discussing the prospect of utilizing them for applications that are likely viable for implementation in the near-term, highlighting benefits and challenges that need to be overcome to execute useful computations.

\subsubsection*{\bf{Multi-cavity quantum processors}}

The U.S. Department of Energy's Superconducting Quantum Materials and Systems (SQMS) center has launched a 3D cavity-based quantum computing program~\cite{alam2022quantum}. Bare SRF cavities at milli-Kelvin temperatures have demonstrated photon lifetimes of $T_1 \sim 2$ seconds for the fundamental mode~\cite{romanenko2020three}, representing a three-orders-of-magnitude improvement over two-dimensional resonators. SQMS center have integrated superconducting transmon qubits with high quality factor $Q$ single-cell Nb SRF cavities, achieving preparation of non-classical cavity states without significantly degrading the cavity's intrinsic lifetime~\cite{roy2024qudit}. 

This development is crucial for qudit-based QPUs, as introducing a transmon qubit into a cavity induces additional loss; the results indicate that by careful engineering, the cavity's coherence time can be in large part preserved and the coupled cavity-qubit system can gain qudit manipulation functionality. Strategies are now being pursued to improve coherence and realize multi-qudit architectures by coupling multiple cavity-transmon modules. On the theoretical side, new frameworks are being developed to guide these hybrid architectures. For instance, \cite{liu2024hybrid} formalizes an instruction set architecture for hybrid cavity–qubit processors, laying out how algorithms can be compiled to operations on coupled transmons and cavities.

The ability to reliably store and manipulate quantum information for extended durations, combined with access to many energy levels within a single mode, motivates the use of cavity resonator modes as qudits~\cite{reineri2023exploration}. A natural encoding for such bosonic qudits is the Fock basis of the cavity modes, the discrete photon-number states $\{|0\rangle, |1\rangle, |2\rangle, \dots, |d-1\rangle\}$, for some arbitrary $d$, which spans a high-dimensional Hilbert space. By restricting to the lowest Fock levels, one can treat a cavity mode as a $d$-level quantum register. Because an isolated harmonic mode has evenly spaced energy levels and no intrinsic nonlinearity, direct control of a cavity's quantum state requires coupling to an ancilla with anharmonic element, such as a superconducting transmon. The transmon is inserted and dispersively coupled to the cavity, so driving the coupled system with carefully chosen microwave pulses can enact sideband transitions exchanging excitations between the qubit and the cavity. For example, a drive at the appropriate sideband frequency can simultaneously lower the transmon from the excited state $|e\rangle$ to the ground $|g\rangle$ while adding a photon to the cavity (transition $|e, n\rangle \to |g, n+1\rangle$), or vice versa. Such operations allow deterministic single-photon creation, annihilation, and Fock-state manipulations. By combining cavity displacements with selective number-dependent arbitrary phase (SNAP) gates – conditional phase shifts on individual Fock states mediated by the transmon – one obtains a universal gateset for a single bosonic mode~\cite{ogunkoya2023investigating}. This control toolkit has experimentally shown high-fidelity preparation of $\simeq$20 Fock states and the implementation of arbitrary unitary transformations, even with relatively weak dispersive coupling~\cite{yaolutalkAPS}. Recent benchmarks indicate that a single transmon can reliably manipulate a cavity qudit spanning tens of photon-number levels with current coherence parameters~\cite{bornman2025benchmarking}.

The same bosonic-mode architecture extends beyond one mode to realize a multi-qudit processor. One approach incorporates multiple long-lived modes within a single 3D environment – for example, using a multi-cell superconducting radio-frequency cavity supporting several resonant modes. Each mode serves as an independent qudit, all interfaced by a common transmon coupler. Through the transmon's nonlinear coupling, interactions between different cavity modes can be activated via virtual Raman processes, wherein simultaneous drives induce effective photon exchange between modes without populating the transmon's excited state. The transmon, with shorter coherence time, is used only as a catalyst for quantum operations – remaining idle except during gate pulses and measurements. These multi-mode operations enable entangling gates between bosonic qudits. Driving the system at the frequency difference of two cavity modes can enact a beam-splitter-like interaction that swaps excitation between modes – entangling their states when properly sequenced with single-mode operations. Two-mode entangling operations have been demonstrated in 3D devices, confirming high-fidelity quantum gates between bosonic modes are achievable. This design minimizes exposure of quantum data to the transmon's decoherence channels, inheriting noise resilience of the bosonic mode while allowing fast, controllable interactions~\cite{roy2024qudit}.

Extrapolating the pace of R\&D progress, and based on disclosed roadmaps by research groups and companies, it's realistic to forecast the feasibility in the near-term of a multi-cell array composed by  $\simeq$10 linearly connected cavities, each contributing $\simeq$4 modes that can be occupied by $d \simeq 10$ photons with millisecond $T_1$ lifetime, within the next 5 years. Such a system would exceed 100 qubits in Hilbert space dimension and could be profitably used for experimentation if certain challenges are solved.

\section{Applications of bosonic qudit computers}

As quantum computing progresses from the NISQ era toward fault-tolerance, recent experiments have demonstrated key milestones in quantum error correction for superconducting cavity architectures. Despite global investment shifting toward reaching the MegaQuOp error-corrected regime~\cite{preskill2025beyond}, NISQ-like experimental campaigns with optimized devices remain invaluable for hardware optimization and discovering effects that could enable quantum advantage. Although conclusive near-term quantum advantage remains elusive, experimentation is an essential strategy in coherent systems with inherently large Hilbert spaces, since their open system dynamics becomes fast intractable for numerical simulations.

Elaborating on ideas recently appearing in the literature, we identify three examples of use-inspired scientific cases that could stress-test these 3D devices' algorithmic capabilities, one for each mainstream application domain: simulation, optimization, and machine learning. In table I we summarize the perspective size and parameters of the discussed applications and the main challenges that need to be overcome.
\subsection{Quantum Simulation: lattice field theory}
One opportunity in the domain of quantum simulation would be to generalize the study in \cite{gustafson2022noise}, which  investigated the simulation of a (1+1)-dimensional U(1) gauge theory with scalar fields (sQED), aiming to extract the mass gap via real-time quantum simulations. After approximations, the Hamiltonian on $N_s$ linear lattice sites can be written in terms of linear and quadratic terms (involving only single or adjacent sites) composed by ladder and diagonal operators, $\hat{L}^z |m\rangle = m |m\rangle$.

These operators theoretically live in an infinite dimensional Hilbert space but must be mapped to finite quantum resources.

In \cite{gustafson2022noise}, qubit and qutrit encodings were compared, mapping infinite-dimensional gauge fields into discrete Z$_3$ qutrit systems. Quantum circuits implemented Trotterized Hamiltonian evolution under realistic noise. Simulations showed that using the most native qutrit encodings tolerated gate errors 10–100 times higher than qubit encodings.

\paragraph*{Identified Opportunity} The Hamiltonian studied in \cite{gustafson2022noise} could be clearly supporting qudits beyond qutrits (max $m$=$d$) and could be generalized on a higher-dimensional lattice. This opens up possibilities for cutting edge examples of 2D simulations by embedding this problem onto a 1D ladder of resonators each supporting two or possibly more bosonic modes so that one could use a generalization of the model above to a pure gauge Hamiltonian of U(1) which has a nearly identical Hamiltonian which is written down in Ref.~\cite{Unmuth-Yockey:2018xak}. This would allow for a first principles real-time demonstration of pure gauge theories in 2+1D on a multi-cavity platform. Going beyond 2D could also be possible for a small number of sites in the near term by expanding the number of addressable mode per cavity and use a swap network to allow 3D interactions.
\paragraph*{Anticipated Challenge} While extrapolations from preliminary numerical results in \cite{gustafson2022noise} are promising, a key challenge is the engineering the CSUM gate, crucial to implement nearest-neighbor interactions. This controlled increment increases the target state by the controlled state value $\text{CSUM} |a\rangle|b\rangle = |a\rangle|b\oplus a\rangle$.
This gate is the Clifford extension of CNOT to qudit states, and efficiently implementing it is key for both near and far-term applications. The timescale of execution of this gate at high fidelity will ultimately determine the viability and scale of the simulation. There has been work on the subject~\cite{PhysRevA.105.042434, mato2023compilation}, but it typically requires advanced pulse and hardware-specific tuning. As this gate forms the basis for entangling operations in the Clifford-basis, it's crucial also for fault-tolerant qudit simulations and represents a missing engineering component beyond sQED simulations.

\subsection{Quantum Optimization: graph coloring}

The NISQ Era has seen many variants of the Quantum Approximate Optimization Algorithm (QAOA) as the baseline method for employing gate-model quantum computers to solve combinatorial optimization problems. Although the variational aspect makes proving utility difficult in general cases, evidence from numerics and theory is starting to appear for scaling advantages that have been long hypothesized~\cite{shaydulin2024evidence, omanakuttan2025threshold}, even considering fault-tolerant overhead~\cite{omanakuttan2025threshold}. These encouraging results remain far from practical. Experiments demonstrate difficulties scaling vanilla QAOA to tens (and recently 100+) qubits without hybridizing with classical HPC and co-designing with hardware specificities.

Another prominent challenge is designing executable circuit dynamics that respect hard constraints between variables~\cite{hadfield2017quantum}, which are ubiquitous in real-world problems. When executing hard-constrained ansatze on noisy QPUs, even with high baseline fidelities, symmetries upholding constraints are quickly destroyed by noise, and the probability of obtaining valid solutions decreases exponentially~\cite{niroula2022constrained}.

Qudits offer an interesting natural mechanism for enforcing one-hot constraints, prominent in scheduling problems. Graph-coloring (i.e. maximizing the number of properly colored edges) demonstrates this encoding~\cite{bravyi2022hybrid}. Colors (corresponding to time-slots in scheduling) are assigned to qudit basis states, with $d$ mapping directly to the number of colors, which can be mixed by single-qudit rotations. The assignment of multiple colors to the same graph node is physically forbidden.

In \cite{ozguler2022numerical}, quantum optimal control techniques were used to numerically synthesize high-fidelity QAOA circuits for bosonic qudit systems implemented via superconducting cavity modes coupled to transmon qubits. Their study demonstrated precise handling of single-qudit rotation operations controlling up to eight energy levels and two-qutrits operations (``phase separation'' in QAOA), achieving gate fidelities exceeding 99\% in noiseless setting.

\paragraph*{Identified Opportunity} 
In combinatorial optimization, we benefit from needing only a single classical bitstring as the final objective, with no requirements on outputs' probability distribution. This allows using dissipative effects as computational search primitives. Recently, the Noise-Directed-Adaptive-Remapping (NDAR) technique successfully used the $Z_2$ gauge degree of freedom in Ising models to map the noise-dominated output of an 84-qubit processor running QAOA to high-quality candidate solutions, dramatically increasing the probability of optimal solutions~\cite{maciejewski2024improving}. This method could generalize to qudits, exploiting photon loss as an asset. 

To scale beyond the variables naturally offered by available modes while preserving constraint-handling capabilities, algorithms could employ quantum-random-access-codes (QRACs)~\cite{fuller2024approximate}, associating variables to multi-body  expectation values. Recently, combinatorial problems with 1000+ nodes were solved using this method on trapped-ion QPUs~\cite{sciorilli2025towards}, though no studies yet generalize these quantum optimization algorithms to qudits.


\paragraph*{Anticipated Challenges}
Similar advances as the ones required for the CSUM gate are required to perform a clear resource estimate for this application as well, but even with imperfect synthesis, hard constraints would be preserved. The numerical studies to synthesize two-qudit gates don't consider decoherence and employ numerical methods that can't scale with increasing Hilbert space. Constructive algorithms for synthesis are the likely solution, yet to be demonstrated in context~\cite{mato2023compilation}.
Despite theory advances and the intellectual merit of investigating QAOA generalizations, quantum methods for optimization will likely realize value only through challenging large-scale experimentation on hard problems using hybrid, novel solver designs.

\begingroup
  \renewcommand\thetable{}%
  \renewcommand\tablename{}%
  
  \begin{table*}[t]
    \caption{Summary of Proposed Application Experiments for next‑gen Superconducting cavity QPU\vspace{-5pt}}
    \label{tab:experiments}  
    \centering
    \begin{tabular}{|c|c|c|}
      \hline
      \textbf{\textit{Application}} 
        & \textbf{\textit{Implementation estimation$^*$}} 
        & \textbf{\textit{Main challenge}} \\
      \hline
      sQED Simulation
        & 2D lattice $N_s=9\times2$ with $d=4+$ following \cite{gustafson2022noise}
        & Synthesis CSUM between co‑located and adjacent qumodes~\cite{job2023efficient}\\
      \hline
      Coloring Optimization
        & NDAR‑QAOA~\cite{maciejewski2024improving} 3‑colors N=9 (or 50+ via QRACS~\cite{sciorilli2025towards})
        & CSUM and generalize QRACs to qudits~\cite{fuller2024approximate}\\
      \hline
      Reservoir Computing
        & time‑series prediction with 1000+ equivalent neurons~\cite{dudas2023quantum}
        & Measurement scheme with low sampling overhead (shot noise)~\cite{wudarski2023hybrid}\\
      \hline
      \multicolumn{3}{l}{%
        $^*$ estimations refer to objectives that are difficult (due to noise)
        but in principle mappable and executable in NISQ hardware.%
      }
    \end{tabular}
    \vspace{-20pt}
  \end{table*}
\endgroup

\subsection{Quantum Machine Learning: reservoir computing}

Quantum Reservoir Computing (QRC) offers a flexible computational framework adaptable to various quantum platforms. Unlike standard quantum neural networks, this paradigm avoids resource-expensive processes like gradient calculation, making it viable for immediate quantum machine learning exploration without encountering barren plateau problems, while demonstrating strong resilience to noise and decoherence. The hybrid transmon-cavity processor could act as a unique \emph{analog} quantum reservoir, as recently demonstrated in two 2024 studies~\cite{senanian2024microwave, krisnanda2024experimental} addressing classical signal processing and quantum state tomography.

In \cite{senanian2024microwave}, analog microwave waveforms are continually fed into the cavity, displacing its state, while the transmon qubit is periodically measured. The coupled qubit–oscillator undergoes entangling unitary evolutions interleaved with measurements, inducing complex nonlinear dynamics that capture temporal correlations of the input signal. The measurements' back-action on the oscillator creates non-unitary evolution, enriching dynamics beyond what a closed system could achieve. Training occurs classically: measurement outcome time-series are linearly combined to classify inputs. Using this analog QRC, the authors successfully distinguished various microwave signal classes with high accuracy, including ultra-low-power signals containing only a few photons.

In \cite{krisnanda2024experimental}, authors demonstrate continuous-variable quantum state and process tomography on a similar setup to reconstruct unknown quantum states of the cavity. During training, known input states are prepared and evolved under a fixed operation sequence in the coupled cavity-qubit system. This involves applying calibrated displacement operations followed by qubit measurements yielding cavity parity information.
A series of displacement-and-measurement steps produces rich measurement outcomes for each input state. Using these as features, a linear model is trained to map the reservoir's measurements to known input state properties. In testing, the system processes unknown states with the same sequence, collects measurement data, and applies the learned linear map to predict the state. A Bayesian inference step enforces physical consistency on the reconstructed density matrix. This approach achieved high-fidelity tomography of various cavity states: the learned reservoir "black-box" automatically compensates for decoherence, control imperfections, and other non-idealities. Notably, the authors observe that this strategy required smaller training datasets and simpler resources than competing methods.

\paragraph*{Identified Opportunity} With emerging multi-mode processors, these experiments performed with a single mode could be excellent quantum reservoirs for experimenting with larger Hilbert spaces. One step in this direction has been numerically studied in \cite{dudas2023quantum} where two bosonic modes were directly coupled to form an interacting reservoir. The reservoir evolves according to the Hamiltonian $H=\sum_i \omega_i a_i^\dagger a_i + g(a_1^\dagger a_2 + h.c)$ where $a_{i=\{1,2\}}$ are the harmonic oscillators, subject to dissipation. The use-case is time-series prediction. Input data is fed by displacing oscillator states or modulating coupling, and outputs are obtained by measuring oscillator observables. This design leverages variable encodings using fock states: with just two oscillators, up to around 9 levels are used to create a reservoir of effectively 81 ``neurons". Achieving similar performance classically required a much larger reservoir and hints at a possible scaling advantage, to be verified by more comprehensive bencharking. Similarly as QRAC encodings, depending on the observable measured, ten oscillators could emulate millions of neurons, in principle~\cite{wudarski2023hybrid}.

\paragraph*{Anticipated Challenges} The first experimental demonstration using multiple oscillators remains to be achieved but is within reach of emerging technology. However, for time-sensitive applications, it will be essential to design measurement schemes that define the input to the trainable classical layer without incurring large shot noise overhead, which quickly degrades performance and would prohibit real-time operation of the system. Moreover, special attention will have to be taken ensuring that the feature map encoding the input won't limit the expressibility of the reservoir~\cite{schutte2025expressive}.
\vspace{-5pt}
\section*{Acknowledgment}
U.S. DOE, Office of Science, NQI Science Research Centers, Superconducting Quantum Materials and Systems Center (SQMS) under Contract DE-AC02-07CH11359.
D.V. and E.G. are under NASA ISRDS-3 Contract 80ARC020D0010.

\bibliographystyle{IEEEtran}
\bibliography{biblio}

\end{document}